\documentstyle[aps,preprint]{revtex}

\begin{document}
\draft

\title
{\bf Combinatorics of Feynman Diagrams for the Problems with Gaussian 
Random Field}
\author{E.Z.Kuchinskii,\ M.V.Sadovskii}
\address
{Institute for Electrophysics, \\
Russian Academy of Sciences,\ Ural Branch,\\
Ekaterinburg 620049, Russia
\\ E-mail: sadovski@ief.intec.ru}
\maketitle

\begin{center}
{\sl Submitted to JETP, May 1997}
\end{center}

\begin{abstract}

The algorithm to calculate the generating function for the number of 
``skeleton'' diagrams for the irreducible self-energy and vertex parts is
derived for the problems with Gaussian random fields. We find an exact
recurrence relation determining the number of diagrams for any given order
of perturbation theory, as well as its asymptotics for the large order limit.
These results are applied to the analysis of the problem of an electron in the
Gaussian random field with the ``white-noise'' correlation function.
Assuming the equality of all ``skeleton'' diagrams for the self-energy part in
the given order of perturbation theory, we construct the closed integral
equation for the one-particle Green's function, with its kernel defined by the
previously introduced generating function. Our analysis demonstrate that this
approximation gives the qualitatively correct form of the localized states
``tail'' in the density of states in the region of negative energies and is
apparently quite satisfactory in the most interesting region of strong
scattering close to the former band-edge, where we can derive the asymptotics
of the Green's function and density of states in the limit of very strong
scattering.

\end{abstract}


\newpage
\narrowtext

\section{Introduction}

Different methods of summation of Feynman diagrams are widely used in rather
wide class of problems of theoretical physics dealing with elementary
excitations propagation in systems with static random fields, created by
different types of inhomogeneites. A simplest example is the problem of an 
electron  in a metal with impurities. Apparently this was the first problem
in which the diagram technique considered in this paper was formulated for 
the first time.\cite{Edw,AGD}. Analogous formalism is being widely used in
problems of statistical radiophysics and optics, dealing with the propagation
of electromagnetic waves in disordered media\cite{RKT}. 
Equivalent mathematical approach is used in a number of problems of critical
phenomena in disordered systems\cite{Ma}, in the problem of polymer chain
with ``excluded volume'' and in some other problems of polymer physics
\cite{GX}. The same diagram technique describes the regular model of critical
phenomena with ``zero-component'' order parameter\cite{Ma}.

In any problem, dealing with summation of Feynman diagrams, any kind of
information on combinatorics of graphs, i.e. on the number of diagrams of
different types for the given order of perturbation theory, is quite useful.
In this paper we present a detailed study of Feynman graphs combinatorics for
the problems with the Gaussian random field.

\section{Generating function for the number of ``skeleton'' diagrams.\ 
Recurrence relation.}

For definiteness we shall always discuss the problem of an electron with
energy $E$ and momentum ${\bf p}$, propagating in the Gaussian random field
(random impurity system) \cite{Edw,AGD}. The averaged Green's function is
determined by the perturbation series shown in Fig.1(a). In a standard way
this series is reduced to Dyson's equation:

\begin{equation}
G(E,p)=\frac{1}{E-\varepsilon_{p} -\Sigma (E,p)}
\label{1}
\end{equation}
where $\varepsilon_{p}=\frac{p^{2}}{2m}$ -- is the free electron spectrum, 
while the self-energy part $\Sigma (E,p)$ is defined by the ``skeleton''
graphs of Fig.1(b), where the internal electronic line represents the fully
renormalized (``dressed'') Green's function $G(E,p)$.

The total number of diagrams in the $N$-th order of perturbation expansion
shown in Fig.1(a) is easily demonstrated to be equal to:

\begin{equation}
G_{N}=(2N-1)!!=\frac{(2N-1)!}{2^{N-1}(N-1)!}
\label{2}
\end{equation}
which is just the total number of ways in which we can connect $2N$ 
vertices by $N$ impurity lines. The analogous number of graphs for
$\Sigma_{N}$ in the expansion of Fig.1(b) is much more difficult to find and
the exact answer for this problem is, as far as we know, absent in the
literature. Only in Ref.\cite{Sus} a simple inequality was shown to hold:

\begin{equation}
(2N-1)!!>\Sigma_{N}>(2N-3)!!
\label{3}
\end{equation}
This inequality gives only a very rough estimate of $\Sigma_{N}$. 
As we shall show below this problem can be solved exactly. This follows
immediately from an exact solution for the electron in the random potential
$V({\bf r})=V$, where the value of $V$ is independent of the spatial
coordinate $\bf r$, though Gaussian with distribution width $<V^{2}>=W^{2}$. 
Naturally, in this case the diagram technique is of the standard form like
in Fig.1, and each impurity line transfers just the zero value of momentum,
i.e. it is associated (in momentum space) with correlator of the form
 $(2\pi )^{d}W^{2}\delta ({\bf q})$ ($d$ -- is spatial dimensionality)
\cite{Kel,ES}. All contributions of the same order in perturbation expansion
of Fig.1(a) are equal and the series for the Green's function can be
expressed as\cite{Kel}:

\begin{equation}
G(E,p)=G_{0}(E,p)\left\{1+
\sum_{N=1}^{\infty}(2N-1)!!G_{0}^{2N}(E,p)W^{2N}\right\}
\label{4}
\end{equation}
Then using the representation:

\begin{equation}
(2N-1)!!=\frac{1}{\sqrt{2\pi}}
\int\limits_{-\infty}^{\infty}dtt^{2N-2}e^{-\frac{t^{2}}{2}}
\label{5}
\end{equation}
the series (\ref{4}) is summed directly and we obtain:
\footnote{Mathematically it is equivalent to Borel summation.}

\begin{equation}
G(E,p)=\frac{1}{W}\Psi\left(\frac{1}{WG_{0}(E,p)}\right)
\label{6}
\end{equation}
where we introduced the function:

\begin{equation}
\Psi (z)=-\frac{1}{\sqrt{2\pi}}
\int\limits_{-\infty}^{\infty}dte^{-\frac{t^{2}}{2}}\frac{1}{t-z}
\label{7}
\end{equation}

Consider the self-energy part corresponding to the Green's function (\ref{6}). 
Addition of a new impurity line to any diagram in this problem leads 
just to the appearance of an additional factor  $W^{2}G^{2}$, and the
self-energy part, defined by the expansion shown in Fig.1(b) can be written 
as:

\begin{equation}
\Sigma = Q(W^{2}G^{2})W^{2}G
\label{8}
\end{equation}
where $Q(x)$ is some function. We shall see, that this function is the
generating function for the number of ``skeleton'' graphs for the
self-energy part, i.e. the coefficients of its Taylor expansion determine
the required numbers $\Sigma_{N}$.

The Dyson's equation for the problem under consideration is written as:

\begin{equation}
G=G_{0}+G_{0}\Sigma G=G_{0}\left(1+Q(W^{2}G^{2})W^{2}G^{2}\right)
\label{9}
\end{equation}
Introducing $z=(WG_{0})^{-1}$ and $y=W^{2}G^{2}$,
from Eqs. (\ref{6}) and (\ref{9}) we obtain the following parametric
representation for $Q(y)$:

\begin{eqnarray}
& & 1+yQ(y)=z\Psi (z)=z\sqrt{y} \nonumber \\
& & y=\Psi ^{2}(z)
\label{10}
\end{eqnarray}
This representation of $Q$ is rather inconvenient. Below we show that this
function obeys certain differential equation. It is easily seen that 
previously introduced function $\Psi (z)$, satisfies the usual dispersion
relation:\footnote{The sign of imaginary part is defined as we consider
either retarded or advanced Green's function.}

\begin{equation}
Re\Psi (z)=\frac{1}{\pi}\int\limits_{-\infty}^{\infty} dt\frac{Im\Psi (t)}{t-z};
\qquad \frac{1}{\pi}Im\Psi (t)=\mp \frac{1}{\sqrt{2\pi}}e^{-\frac{t^{2}}{2}}
\label{11}
\end{equation}
from which it follows immediately that $\Psi (z)$ satisfies the differential
equation: 

\begin{equation}
\frac{d\Psi}{dz}=1-z\Psi
\label{12}
\end{equation}
with the initial condition:
\begin{equation}
\Psi (z=\pm i0)=\mp i\sqrt{\frac{\pi}{2}}
\label{13}
\end{equation}

Differentiating the first equation in (\ref{10}) over $y$, we obtain:

\begin{equation}
\frac{dz}{dy}=\frac{1}{2}y^{-\frac{3}{2}}
\left\{2y^{2}\frac{dQ(y)}{dy}+yQ(y)-1\right\}
\label{14}
\end{equation}
Differentiating the second-equation in (\ref{10}) over $z$ and using 
(\ref{12}) we obtain:

\begin{equation}
\frac{dy}{dz}=2\Psi (z)\frac{d\Psi (z)}{dz}=2\Psi (z)(1-z\Psi (z))
=-2y^{\frac{3}{2}}Q(y)
\label{15}
\end{equation}
Comparing (\ref{14}) and (\ref{15}) we obtain the following non-linear 
differential equation for $Q(y)$:

\begin{equation}
\frac{dQ(y)}{dy}=\frac{1}{2y^{2}}\left\{1-Q^{-1}(y)+yQ(y)\right\}
\label{16}
\end{equation}
Using (\ref{10}) and (\ref{13}) we get $y=\left.\Psi^{2}(z)\right|_{z=\pm 
i0} =-\frac{\pi}{2}$, so that

\begin{equation}
Q(-\frac{\pi}{2})=\left.\frac{z\Psi (z)-1}{y}\right|_{z=\pm i0} =\frac{2}{\pi}
\label{17}
\end{equation}
which gives the initial condition for Eq.(\ref{16}). Note that the condition
$Q(0)=1$, which obviously follows from diagram expansion for $\Sigma$,
is a special one for Eq.(\ref{16}) and can not be used as an initial
condition.

Eq.(\ref{16}) can be expressed in more convenient form:

\begin{equation}
Q(y)=1+y\frac{d}{dy}yQ^{2}(y)
\label{18}
\end{equation}
We are interested in Taylor expansion for $Q(y)$:

\begin{equation}
Q(y)=\sum_{n=0}^{\infty}a_{n}y^{n}
\label{19}
\end{equation}
As the number of ``skeleton'' diagrams of the  $N$-th order for the
self-energy part is just the coefficient of $W^{2N}$ in the expansion of
$\Sigma$ over the powers of $W^{2}$, it is easily seen that Eq.(\ref{8}) 
gives the required value of $\Sigma_{N}$ as:

\begin{equation}
\Sigma_{N}=a_{N-1}
\label{20}
\end{equation}
This means that the function $Q(y)$ is the generating function for
combinatorial factors $\Sigma_{N}$.

Substitution of Eq.(\ref{19}) into Eq.(\ref{18}) leads to the following
recurrence relation for the coefficients $a_{n}$:

\begin{equation}
a_{n}=n\sum_{m=0}^{n-1}a_{m}a_{n-1-m}
\label{21}
\end{equation}
where $a_{0}=1$. From $a_{0}=1$ it follows that $Q(0)=1$. It is due to this
fact this point is special -- equation $Q(0)=1$ is satisfied for any initial
conditions, for which Eq.(\ref{18}) has a solution.

From Eq.(\ref{21}) it is easy to find the values of $a_{n}$ for small 
$n$, the appropriate results are presented in Table 1.

The knowledge of combinatorics for the self-energy part allows to find also
the combinatorics for the two-particle Green's function -- both for the
full vertex-part $\Gamma$ and for the irreducible vertex $U$. The appropriate
diagram representations of these vertices is shown in Fig.2. 
The self-energy part $\Sigma$ is connected with the vertex-part $\Gamma$ 
by the equation shown graphically in Fig.3. For the problem with zero
transferred momentum\cite{Kel,ES} this equation has the following form:

\begin{equation}
\Sigma =W^{2}G(1+G^{2}\Gamma )
\label{22}
\end{equation}
Thus, for the number of diagrams in the  $N$-th order of the full vertex
$\Gamma_{N}$, we obtain immediately:

\begin{equation}
\Gamma_{N}=\Sigma_{N+1}=a_{N}
\label{23}
\end{equation}
In this sense  $Q(y)$ is also the generating function for the number of
diagrams for the full vertex-part.

The number of diagrams of the $N$-th order for the irreducible vertex-part
$U_{N}$ can be easily obtained if we note, that the cut of any of  $2N-1$ 
internal Green's function lines in the diagram for the self-energy part of
$N$-th order produces the appropriate diagram for the $N$-th order
contribution to $U$ (Fig.4).  Thus:

\begin{equation}
U_{N}=(2N-1)\Sigma_{N}=(2N-1)a_{N-1}
\label{24}
\end{equation}

In Appendix A we once again derive the differential equation (\ref{18}) for
the generating function $Q(y)$, using only the Bethe-Salpeter equation,
connecting $U$ and $\Gamma$, as well as the Ward-type identity, with no use
of the explicit solution for the Green's function (\ref{6}).

\section{Asymtotics for the number of diagrams in the limit of large $N$.}

In the limit of large orders of perturbation theory $N\gg 1$ the use of the
recurrence relation (\ref{21}) becomes inconvenient due to the factorial
growth of the number of diagrams\cite{Sus}. At the same time the fact of
this factorial growth itself can be used for significant simplification of
the problem. Let us rewrite (\ref{21}) as: 

\begin{equation}
a_{n}=2na_{0}a_{n-1}+2na_{1}a_{n-2}+2na_{2}a_{n-3}+\cdots
\label{25}
\end{equation}
where $a_{0}=1$, $a_{1}=1$, $a_{2}=4$. It is natural to assume that in the
limit of large $n$ we have $a_{n}\approx (2n+\beta )a_{n-1}$, then 
$a_{n-2}\approx \frac{a_{n-1}}{2n-2+\beta }$ etc. Substituting these
expressions into (\ref{25}) immediately leads to $\beta =1$ and 

\begin{equation}
a_{n}=(2n+1+O(\frac{1}{n}))a_{n-1}
\label{26}
\end{equation}
This means that in the limit of large $n$ we have $a_{n}\sim (2n+1)!!$.
Let us define $b_{n}$ as:

\begin{equation}
b_{n}=\frac{a_{n}}{(2n+1)!!}
\label{27}
\end{equation}
Substituting (\ref{27}) into (\ref{21}), we obtain the recurrence relation
for $b_{n}$:

\begin{equation}
b_{n}=n\sum_{m=0}^{n-1}\frac{(2m+1)!!(2n-2m-1)!!}{(2n+1)!!}b_{m}b_{n-1-m}
\label{28}
\end{equation}
with $b_{0}=1$. In the limit of large $n$, taking into account
$b_{1}=\frac{1}{3}$, $b_{2}=\frac{4}{15}$, and limiting ourselves to the
accuracy of the order of $b/n^{2}$ 
(where $b\sim b_{n}\sim b_{n-1}\sim b_{n-2}\sim b_{n-3}$), we get:

\begin{equation}
\Delta b_{n}=b_{n}-b_{n-1}=\frac{5}{4}\frac{b_{n-1}}{n^{2}}+O(\frac{b}{n^{3}})
\label{29}
\end{equation}
Thus, in the limit of large $n$ we can write down the following differential
equation for $b_{n}$:

\begin{equation}
\frac{db_{n}}{dn}=\frac{5}{4}\frac{b_{n}}{n^{2}}+O(\frac{b}{n^{3}})
\label{30}
\end{equation}
from which it follows immediately:

\begin{equation}
b_{n}=b\cdot exp\left(-\frac{5}{4}\frac{1}{n}+O(\frac{1}{n^{2}})\right)=
b\left\{1-\frac{5}{4}\frac{1}{n}+O(\frac{1}{n^{2}})\right\}
\label{31}
\end{equation}
It is natural, that this analysis can not provide us with the value of the
constant $b=\lim_{n \to \infty}b_{n}$. Numerical study of $b_{n}$ using the
recurrence relation (\ref{28}) completely supports the dependence defined by
Eq.(\ref{31}) (Cf.Fig.5) and leads to the value of 
$b=\frac{1}{e}=0.36787944\cdots$ 
(calculations were done up to $n=5000$, which guarantees the claimed 
accuracy). We do not know any analytical way to obtain this rather curious
result.

Finally, the asymptotic expressions for the number of diagrams of different
types for large $N$ have the following form: 
\footnote{Asymptotic dependence of the type of Eq.(\ref{32})
$\Sigma_{N}\approx c2^{N}\Gamma (N+\beta )$ was obtained in Ref.\cite{Sus}.
However, the values of the coefficients 
$c$ and $\beta$ were not determined.}

\begin{eqnarray}
\Sigma_{N}&=&a_{N-1}=b_{N-1}(2N-1)!!=\frac{1}{e}\left\{1-\frac{5}{4}\frac{1}{N}+O(\frac{1}{N^{2}})\right\}(2N-1)!!= \nonumber \\
&=&\frac{1}{\sqrt{\pi}e}\left\{1-\frac{5}{4}\frac{1}{N}+O(\frac{1}{N^{2}})\right\}2^{N}\Gamma (N+\frac{1}{2})
\label{32}
\end{eqnarray}

\begin{eqnarray}
\Gamma_{N}&=&a_{N}=\frac{1}{e}\left\{1-\frac{5}{4}\frac{1}{N}+O(\frac{1}{N^{2}})\right\}(2N+1)!!= \nonumber \\
&=&\frac{1}{\sqrt{\pi}e}\left\{1-\frac{5}{4}\frac{1}{N}+O(\frac{1}{N^{2}})\right\}2^{N+1}\Gamma (N+\frac{3}{2})
\label{33}
\end{eqnarray}

\begin{eqnarray}
U_{N}&=&(2N-1)a_{N-1}=\frac{1}{e}\left\{1-\frac{5}{4}\frac{1}{N}+O(\frac{1}{N^{2}})\right\}(2N-1)(2N-1)!!= \nonumber \\
&=&\frac{1}{e}\left\{1-\frac{9}{4}\frac{1}{N}+O(\frac{1}{N^{2}})\right\}(2N+1)!!=
\frac{1}{\sqrt{\pi}e}\left\{1-\frac{9}{4}\frac{1}{N}+O(\frac{1}{N^{2}})\right\}2^{N+1}\Gamma (N+\frac{3}{2})  
\label{34}
\end{eqnarray}
It is interesting to note that:

\begin{equation}
\frac{\Sigma_{N}}{G_{N}}=b_{N-1}=
\frac{1}{e}\left\{1-\frac{5}{4}\frac{1}{N}+O(\frac{1}{N^{2}})\right\}
\rightarrow \frac{1}{e}
\label{35}
\end{equation}

\begin{equation}
\frac{U_{N}}{\Gamma_{N}}=1-\frac{1}{N}+O(\frac{1}{N^{2}})\rightarrow 1
\label{36}
\end{equation}
In Table 1 we present the summary of the main results for the number of
diagrams of different types.

\section{Electron in the Gaussian random field with the ``white-noise''
correlator.}

As an example of the practical use of the results obtained above, let us
consider the problem of an electron in the Gaussian random field with
``white-noise'' correlator, when the expression associated with impurity
interaction line is\cite{Edw,AGD,MVS}:

\begin{equation}
w({\bf p}_{1},{\bf p}_{2},{\bf p}_{3},{\bf p}_{4})=
W^{2}\delta ({\bf p}_{1}-{\bf p}_{2}+{\bf p}_{3}-{\bf p}_{4})
\label{37}
\end{equation}
where $W^{2}=\rho V^{2}$, $\rho$ -- density of impurities, $V$ -- 
Born scattering amplitude of the point-like impurity. It is well known that
the basic difficulties in this problem appear in the energy region, defined
by the condition\cite{MVS}:

\begin{equation}
|E|\stackrel{<}{\sim}\gamma (E)\text{ или }|E|\stackrel{<}{\sim}E_{sc}
\label{38}
\end{equation}
where $\gamma (E)=\pi\rho V^{2}N(E)$ -- is Born scattering rate
($N(E)$ -- density of states at the energy $E$), $E_{sc}\sim 
m^{\frac{d}{4-d}}(\rho V^{2})^{\frac{2}{4-d}}$ -- characteristic size of
the ``critical'' region around the band-edge, where the strong scattering
appears. These difficulties are mainly due to the impossibility to sum any
kind of dominating series of Feynman diagrams, analogous to that being
summed in the limit of weak scattering
$E\gg \gamma (E)$, $E\gg E_{sc}$ \cite{Edw,AGD}
\footnote{In this case diagrams with noncrossing impurity lines dominate and
we have only to take into account the first diagram in Fig.1(b)}. 
In fact all the diagrams for the self-energy part become of the same order
of magnitude in the energy region $|E|\stackrel{<}{\sim}E_{sc}$ and should be
taken into account.

Perturbation expansion for the self-energy part in terms of ``skeleton''
graphs is shown in Fig.1(b). All the graphs of 3-rd order in this expansion
are in fact equal to each other (Appendix B). Despite the fact that this
property is lost already in the next order, it seems reasonable to
formulate an approximation {\em assuming the equality} of all diagrams of
this type in every order of perturbation theory. Apparently this 
approximation can be good enough especially in the ``critical region''
$|E|\stackrel{<}{\sim}E_{sc}$ where all contributions are of the same order
of magnitude. Let us take as a ``basic'' graph in any given order the
``maximally crossed'' diagram like shown in Fig.6(a). For systems invariant
to time reversal this graph can be transformed into a ``ladder''--like, as
shown in Fig.6(b). Then the full series for the self-energy part in our
approximation can be expressed like:

\begin{eqnarray}
\Sigma (p)&=&\sum_{n=1}^{\infty}W^{2}\Sigma_{n}\sum_{{\bf p}_{1}}\sum_{{\bf p}_{2}}
\left[W^{2}G({\bf p}_{1}+{\bf p}_{2}+{\bf p})G(-{\bf p}_{2})\right]^{n-1}G({\bf p}_{1})= 
\nonumber \\
&=&\sum_{{\bf p}_{1}}W^{2}
Q\left[W^{2}\sum_{{\bf p}_{2}}G({\bf p}_{1}-{\bf p}_{2}+{\bf p})G({\bf p}_{2})\right]G({\bf p}_{1})
\label{39}
\end{eqnarray}
where we used notations defined in (\ref{19}) and (\ref{20}), as well as the
property $G({\bf p})=G(-{\bf p})$, valid for the isotropic system.
Accordingly we obtain {\em closed} equation for the averaged one-particle
Green's function: 

\begin{equation}
G^{-1}(p)=G_{0}^{-1}(p)-W^{2}\sum_{\bf q}
Q\left[W^{2}\sum_{{\bf p}_{1}}G({\bf p}_{1}-{\bf q})G({\bf p}_{1})\right]G({\bf p}+{\bf q})
\label{40}
\end{equation}
where $G_{0}^{-1}(p)=E-\frac{p^{2}}{2m}$. All the non-trivial part of the
problem is contained now in our generating function $Q(y)$, which defines
the ``kernel'' of the complicated nonlinear integral equation (\ref{40}). 
Naturally, if we limit ourselves to the first term in the expansion 
(\ref{19}), we get  $Q=1$ and Eq.(\ref{40}) reduces to the standard sum
of ``noncrossing'' diagrams\cite{Edw,AGD}. The obvious advantage of 
Eq.(\ref{40}) in comparison with standard approach\cite{Edw,AGD}, based upon
the summation of dominating diagrams (e.g. accounting only the first graph
of Fig.1(b)) is the formal account of {\em all} diagrams, which is made,
however, in the approximation, assuming the equality of all the ``skeleton''
graphs for the self-energy part in a given order of perturbation theory.

Equation (\ref{40}) is quite complicated nonlinear integral equation and
can not be solved in general case, more so due to the fact that we do not
know the general form of $Q(y)$ (which enters Eq.(\ref{40}) as a function of
a complex argument). Below we shall limit ourselves with some qualitative
analysis of Eq.(\ref{40}). Let us write Eq.(\ref{40}) in the following
compact form:

\begin{equation}
G^{-1}(p)=G_{0}^{-1}(p)-W^{2}
Q\left[W^{2}G\otimes G\right]\otimes G
\label{41}
\end{equation}
where we have introduced the generalized product (convolution) of functions 
as:

\begin{equation}
F\otimes \Phi =\sum_{\bf p}F({\bf p}-{\bf q})\Phi({\bf p})
\label{42}
\end{equation}
Let us return to Eqs.(\ref{10}), defining  $Q$ parametrically.
The second equation in (\ref{10}) can be written now as:

\begin{equation}
G\otimes G=\frac{1}{W^{2}}\Psi ^{2}(z)
\label{43}
\end{equation}
We have seen that in the problem with zero transferred momentum
$z=W^{-1}G_{0}^{-1}$. Consider now the limit of $W\rightarrow 0$ in 
(\ref{43}). Then the left-hand side of (\ref{43}) reduces to 
$G_{0}\otimes G_{0}$, while in the right-hand side we can assume
$z\sim W^{-1}$ in analogy with the problem with zero transferred momentum and
use the easily established asymptotics $\Psi (z)\approx \frac{1}{z}$ 
for $|z|\gg 1$. Here we are slightly inaccurate, since the exact form of 
$\Psi (z)$ is:

\begin{equation}
\Psi (z)=R(z)\mp i\sqrt{\frac{\pi}{2}}e^{-\frac{z^{2}}{2}}
\label{44}
\end{equation}
where for $R(z)$ we have asymptotic expansion:

\begin{equation}
R(z)=e^{-\frac{z^{2}}{2}}\int\limits_{0}^{z}e^{\frac{t^{2}}{2}}dt=
\frac{1}{z}+\frac{1}{z^{3}}+\frac{3}{z^{5}}+\cdots 
\qquad(-\frac{\pi}{4}<argz<\frac{\pi}{4})
\label{45}
\end{equation}
We are using the asymptotics $\Psi (z)\approx \frac{1}{z}$, which is not
rigorous, however, the results obtained below using this approximation
are confirmed by more accurate, but rather long analysis. Thus, in the limit
of $W\rightarrow 0$ Eq.(\ref{43}) reduces to:

\begin{equation}
G_{0}\otimes G_{0}=\frac{1}{W^{2}z^{2}}\text{ или }z=
\frac{1+O(W^{2})}{W\sqrt{G_{0}\otimes G_{0}}}
\label{46}
\end{equation}
Accordingly, in the limit of $W\rightarrow 0$ instead of (\ref{43}) we can
write:

\begin{equation}
G\otimes G=\frac{1}{W^{2}}
\Psi ^{2}\left(\frac{1}{W\sqrt{G_{0}\otimes G_{0}}}\right)
\label{47}
\end{equation}
Consider the energy region $E<0$, where the fluctuation ``tail'' in the
density of states appears\cite{MVS,LGP}. In this case from (\ref{46}) 
we obviously have $z\in Re$. With the help of (\ref{44}) and (\ref{46}) 
we obtain from (\ref{47}):

\begin{equation}
G\otimes G\approx G_{0}\otimes G_{0}-i\frac{2}{W}\sqrt{\frac{\pi}{2}}
\sqrt{G_{0}\otimes G_{0}}exp\left\{-\frac{1}{2W^{2}G_{0}\otimes G_{0}}\right\}
\label{48}
\end{equation}
where the second term, as we shall see now, produces the fluctuation ``tail''
in the density of states. using $\sum_{\bf q}G\otimes G=
\sum_{\bf p}\sum_{\bf q}G({\bf p}-{\bf q})G({\bf p})=
\left(\sum_{\bf p}G(p)\right)^{2}$, we immediately obtain from (\ref{48})
the density of states in the form:

\begin{equation}
N(E)=-\frac{1}{\pi}\sum_{\bf p}ImG^{R}(E,p)=\frac{1}{\sqrt{2\pi}W}
\frac{\sum_{\bf q}\sqrt{G_{0}\otimes G_{0}}exp\left\{-\frac{1}{2W^{2}G_{0}\otimes G_{0}}\right\}}
{\left|\sum_{\bf p}G_{0}(E,p)\right|}
\label{49}
\end{equation}
Now everything is defined by the concrete form of $G_{0}\otimes G_{0}$ 
for different spatial dimensions. The denominator of (\ref{49}) is easily
calculated in general case as: 

\begin{equation}
\left|\sum_{\bf p}G_{0}(E,p)\right|=
S_{d}\int\limits_{0}^{p_{0}}dpp^{d-1}\frac{1}{|E|+\frac{p^{2}}{2m}}
\label{50}
\end{equation}
where $S_{d}=2^{-(d-1)}\pi ^{d/2}\frac{1}{\Gamma (d/2)}$ while the value of
$p_{0}$ -- represents the cutoff of the order of inverse interatomic
distance\cite{MVS}, which is necessary for $d\geq 2$ 
(for $d=1$ in (\ref{50}) we can extend integration up to infinity).
Now $E$ in Eq.(\ref{50}) denotes the renormalized energy, calculated with
respect to the  band-edge, determined in ``one-loop'' approximation\cite{MVS}, 
which takes into account the infinite (in the limit of 
$p_{0}\rightarrow\infty$ (for $d\geq 2$)) shift of this edge. The integral in
(\ref{50}) is easily calculated for any value of $d$.

In one-dimensional ($d=1$) case all the integrals, contributing to  (\ref{49}), 
are calculated exactly. After rather tedious, but elementary, calculations
we obtain (Appendix C):

\begin{equation}
N(E)=\frac{1}{2\pi}\sqrt{\frac{2m}{|E|}}
exp\left\{-\sqrt{2}\frac{|E|^{\frac{3}{2}}}{m^{\frac{1}{2}}W^{2}}\right\}
\label{51}
\end{equation}
The exponential factor in  (\ref{51}) differs from the well known exact
result of Halperin\cite{Halp} (Cf. also Ch.II in \cite{LGP}) by the absence
of the factor of $\frac{4}{3}$. Preexponential factor in (\ref{51}) is also
different from the exact value which is of the order of
$\sim\frac{|E|}{W^{2}}$ \cite{Halp}. However, the qualitative form of the
``tail'' in the density of states is reproduced more or less satisfactorily,
despite rather common opinion, that this ``tail'' can not be obtained using
the usual perturbation theory.

Similar, but approximate, calculation of the density of states via Eq.(\ref{49})
fro $d=3$ (Appendix D) gives:

\begin{equation}
N(E)=\frac{\pi ^{2}}{12}(2m)^{\frac{3}{4}}\frac{|E|^{\frac{5}{4}}}{E_{0}^{\frac{1}{2}}}\frac{1}{W}
exp\left\{-\sqrt{2}\pi\frac{|E|^{\frac{1}{2}}}{m^{\frac{3}{2}}W^{2}}\right\}
\label{52}
\end{equation}
where we have introduced the cutoff energy 
$E_{0}=\frac{p_{0}^{2}}{2m}$ \cite{MVS}. Here again the exponential factor
coincides (up to a constant) with known results of non-perturbative instanton
approach\cite{MVS,Card,Sad79,Sus97}, while the preexponential in (\ref{52}) 
does not agree with any of the variants obtained in these papers. However,
in general, the result of Eq.(\ref{52}) is again satisfactory enough, taking
into account the approximate nature of our Eq.(\ref{40}).
\footnote{For $d>4$ the use of asymptotic expression of Eq.(\ref{32}) and
statistical analysis of Ref.\cite{Sus} allows to find the {\em correct} power
of $W^{-1}$ in the preexponential factor of the density of states. In this
case our approximation is equivalent to the hypothesis of stationarity of
higher-order contributions used in Ref.\cite{Sus}, which is valid for $d>4$.}

Especially interesting is the analysis of Eq.(\ref{40}) in the 
``strong coupling'' region\cite{MVS}, which is defined by (\ref{38}), 
i.e. in the vicinity of the initial band-edge, where the transition from
extended to localized states takes place. In this region it is reasonable
to assume, that the approximation of equal ``skeleton'' graphs contributions
to the self-energy in the given order of perturbation theory can be rather
good, simply due to the known fact that they are of the same order of
magnitude. Strong condition of the type of Eq.(\ref{38}) is obviously
equivalent to the limit of $W\rightarrow\infty$. In this limit, in 
``zero-order'' approximation we can neglect the first term in the r.h.s. of
Eq.(\ref{41}) and write:

\begin{equation}
G^{-1}(p)=-W^{2}Q\left[W^{2}G\otimes G\right]\otimes G
\label{53}
\end{equation}
It is easy to convince oneself that this is equivalent to the limit of
$z=\pm i0$ in (\ref{43}) or $y=-\frac{\pi}{2}$ in ({\ref{10}). In this case
(\ref{43}) reduces to:

\begin{equation}
W^{2}G\otimes G=\Psi (z=\pm i0)=-\frac{\pi}{2}
\label{54}
\end{equation}
and from (\ref{17}) we get:

\begin{equation}
Q\left[W^{2}G\otimes G\right]=\frac{2}{\pi}
\label{55}
\end{equation}
Formal solution of Eq.(\ref{54}) has the form:

\begin{equation}
G=\pm i\sqrt{\frac{\pi}{2}}\frac{1}{W\sqrt{\aleph}}
\label{56}
\end{equation}
where $\aleph=\sum_{\bf p}1$ -- is the number of states in the band.
Direct substitution of (\ref{56}) and (\ref{55}) into (\ref{53}) shows that
this equation is satisfied. Thus in the ``first-order'' approximation, in the
limit of $W\rightarrow\infty$ we can write down the Green's function
(\ref{41}) as:

\begin{equation}
G(p)=\frac{1}{G_{0}^{-1}(p)-\frac{2}{\pi}W^{2}\sum_{\bf p}G(p)}
\label{57}
\end{equation}
which is surprisingly coincides with the result of the self-consistent
Born approximation (equivalent to the first diagram in Fig.1(b) or Fig.3)
\cite{Edw,AGD}, only with an extra factor of $\frac{2}{\pi}$. 
Obviously Eq.(\ref{57}) leads to the density of states of the Born 
approximation $N_{0}(E)$, which practically coincides for $d=3$ with the
density of states of free electrons (with the account of the shift of the
band-edge in one-loop approximation). In Fig.7 we compare the results
following from Eq.(\ref{57}) for the density of states in one-dimensional
($d=1$) system with the exact result of Halperin\cite{Halp}, which
demonstrates rather satisfactory agreement in the region of ``strong 
coupling'' $|E|<E_{sc}\sim m^{\frac{1}{3}}W^{\frac{4}{3}}$. The width of this
region grows with the growth of $W$. Note that while the ``tail'' of the
density of states is suppressed with the growth of $W$ (Cf.(\ref{51})), 
the crossover region, where $|E|\sim E_{sc}$, becomes wider.

It is possible, that Eq.(\ref{57}) provides the qualitative justification of
the use of the simplest Born--like approximation for the Green's function in
the approaches similar to the self-consistent theory of localization
\cite{MVS,VW} -- the mobility edge appears in the ``strong coupling'' region
$|E|\stackrel{<}{\sim}E_{sc}$ (\ref{38}), where the approximations leading to
(\ref{57}) becomes rather satisfactory and the Green's function really takes
the simple Born--like form.

These results demonstrate the effectiveness of the use of diagram
combinatorics in formulation of new approximations for realistic physical
problems.

This work is partially supported by the Russian Foundation for Basic Research
under the grant No.96-02-16065, as well as from the Project No.IX.1 of the
State Program ``Statistical Physics'' of the Russian Ministry of Science.
The authors are grateful to Dr. A.I.Posazhennikova for the help with
numerical calculations.

\newpage

\appendix
\section{}

Let us derive Eq.(\ref{18}) for generating function $Q(y)$ without the use of
the explicit form of one-particle Green's function (\ref{6}). 
In the problem with zero transferred momentum the Bethe-Salpeter equation
shown in Fug.2(c) reduces to:

\begin{equation}
\Gamma =U+UG^{2}\Gamma
\label{a1}
\end{equation}
so that
\begin{equation}
\Gamma =\frac{U}{1-UG^{2}}
\label{a2}
\end{equation}
Using (\ref{a2}) and (\ref{22}) we obtain an equation, connecting the
self-energy part with irreducible vertex $U$:

\begin{equation}
\Sigma =\frac{W^{2}G}{1-UG^{2}}
\label{a3}
\end{equation}
Use now the  ``Ward identity'':

\begin{equation}
W^{2}\left.\frac{\partial}{\partial W}\right|_{G}\frac{\Sigma}{W}=UG
\label{a4}
\end{equation}
which can be derived via (\ref{8}) and (\ref{24}), 
and Eq.(\ref{a3}) to obtain: 
$$W^{2}\left.\frac{\partial}{\partial W}\right|_{G}\frac{\Sigma}{W}=UG=
\frac{1}{G}\left\{1-W^{2}\frac{G}{\Sigma}\right\}$$

\begin{equation}
\text{или }\qquad \Sigma=W^{2}G+
W^{2}G\Sigma \left.\frac{\partial}{\partial W}\right|_{G}\frac{\Sigma}{W}
\label{a5}
\end{equation}
Now from (\ref{8}), we obtain our differential equation for $Q$:
$$Q(W^{2}G^{2})=1+W^{2}GQ(W^{2}G^{2})\left.\frac{\partial}{\partial W}\right|_{G}
WGQ(W^{2}G^{2})=$$ 
$$=1+W^{2}G^{2}\frac{d}{d(W^{2}G^{2})}W^{2}G^{2}Q^{2}(W^{2}G^{2})$$
which can be rewritten as:

\begin{equation}
Q(y)=1+y\frac{d}{dy}yQ^{2}(y)
\label{a6}
\end{equation}
Note, however, that from this analysis it is impossible to find the correct
initial condition (\ref{17}), which is closely related to (\ref{11}), 
reflecting the causality principle.

\newpage
\section{}

Let us show that all diagrams of the 3-rd order, shown in Fig.1(b) are
equal. Consider the ``maximally crossed'' graph of Fig.1(b.1).  
Its analytical form is:

\begin{eqnarray}
\text{Fig.1(b.1)}&=&W^{6}\sum_{{\bf p}_{1},{\bf p}_{2},{\bf p}_{3},{\bf p}_{4},{\bf p}_{5}}
G(p_{1})G(p_{2})G(p_{3})G(p_{4})G(p_{5})\times \nonumber \\
&\times &\delta ({\bf p}-{\bf p}_{1}+{\bf p}_{3}-{\bf p}_{4})
\delta ({\bf p}_{1}-{\bf p}_{2}+{\bf p}_{4}-{\bf p}_{5})
\delta ({\bf p}_{2}-{\bf p}_{3}+{\bf p}_{5}-{\bf p})
\label{b1}
\end{eqnarray}
In isotropic system the Green's function depends only on the absolute 
magnitude of the momentum and $G({\bf p})=G(-{\bf p})$. 
In (\ref{b1}) we can make the following change of variables of integration
(which obviously does not change the value of integrals): 
${\bf p}_{1}\rightleftharpoons -{\bf p}_{3}$; 
${\bf p}_{2}\rightleftharpoons -{\bf p}_{2}$.
Correspondingly:
$$\delta ({\bf p}-{\bf p}_{1}+{\bf p}_{3}-{\bf p}_{4})
\delta ({\bf p}_{1}-{\bf p}_{2}+{\bf p}_{4}-{\bf p}_{5})
\delta ({\bf p}_{2}-{\bf p}_{3}+{\bf p}_{5}-{\bf p})\rightarrow $$

\begin{equation}
\rightarrow \delta ({\bf p}-{\bf p}_{1}+{\bf p}_{3}-{\bf p}_{4})
\delta ({\bf p}_{2}-{\bf p}_{3}+{\bf p}_{4}-{\bf p}_{5})
\delta ({\bf p}_{1}-{\bf p}_{2}+{\bf p}_{5}-{\bf p})
\label{b2}
\end{equation}
and we obtain the contribution of diagram shown in Fig.1(b.2).
Analogously, the change of variables
${\bf p}_{2}\rightleftharpoons -{\bf p}_{4}$; 
${\bf p}_{3}\rightleftharpoons -{\bf p}_{3}$
produces the contribution of diagram of Fig.1(b.4), while
${\bf p}_{3}\rightleftharpoons -{\bf p}_{5}$; 
${\bf p}_{4}\rightleftharpoons -{\bf p}_{4}$
gives the contribution of Fig.1(b.3). Thus, all the contributions of
``skeleton'' diagrams of the 3-rd order are equal to each other.

Analogous change of integration variables allows to show that under any
impurity in the given ``skeleton'' diagram we can actually  ``rotate''
the electronic line with all incoming impurity lines, so that the new
diagram will give the same contribution as an initial one. This symmetry
produces rather wide classes of equal diagrams in any order of perturbation
theory.

\newpage
\section{}

Here we present some details of calculations, leading from (\ref{49}) 
to (\ref{51}) for one-dimensional ($d=1$) case. For $E<0$ the convolution of
Green's functions, entering (\ref{49}) is calculated by usual contour
integration:

\begin{equation}
G_{0}\otimes G_{0}=\frac{1}{2\pi}\int\limits_{-\infty}^{\infty}dp
\frac{1}{|E|+p^{2}-2pq+q^{2}}\frac{1}{|E|+p^{2}}=
\frac{1}{\sqrt{|E|}(q^{2}+4|E|)}
\label{c1}
\end{equation}
(Here and below for brevity we put $2m=1$.) Then:

\begin{eqnarray}
\sum_{\bf q}\sqrt{G_{0}\otimes G_{0}}exp\left\{-\frac{1}{2W^{2}G_{0}\otimes G_{0}}\right\}
&=&\frac{1}{\pi |E|^{\frac{1}{4}}}\int\limits_{0}^{\infty}\frac{dq}{\sqrt{q^{2}+4|E|}}
exp\left\{-\frac{\sqrt{|E|}}{2W^{2}}(q^{2}+4|E|)\right\}=\nonumber \\
=\frac{1}{2\pi |E|^{\frac{1}{4}}}exp\left\{-\frac{|E|^{\frac{3}{2}}}{W^{2}}\right\}
K_{0}\left(\frac{|E|^{\frac{3}{2}}}{W^{2}}\right)&=&
\frac{W}{2\sqrt{2\pi}|E|}exp\left\{-\frac{2|E|^{\frac{3}{2}}}{W^{2}}\right\}
\label{c2}
\end{eqnarray}
After that by elementary calculations we get:
\begin{equation}
\left|\sum_{\bf p}G_{0}(p)\right|=
\frac{1}{\pi}\int\limits_{0}^{\infty}dp\frac{1}{|E|+p^{2}}=\frac{1}{2\sqrt{|E|}}
\label{c3}
\end{equation}
Putting all these expressions into (\ref{49}) and restoring  $2m$, 
we obtain (\ref{51}).

\newpage
\section{}

Here we present the details of calculations leading from (\ref{49}) to 
(\ref{52}) for $d=3$. Calculations in this case are rather tedious and it is
convenient to use the coordinate representation for the Green's functions.
In the region of $E<0$ we finally obtain:

\begin{equation}
G_{0}\otimes G_{0}=\left\{\begin{array}{ll}
\frac{1}{8\pi\sqrt{|E|}}&\qquad q\ll 2\sqrt{|E|} \\
\frac{1}{8q}&\qquad q\gg 2\sqrt{|E|}
\end{array}\right.
\label{d1}
\end{equation}
In inequalities in  (\ref{d1}) it is convenient to replace $2$ by $\pi$, 
which guarantees the smooth crossover between two limiting expressions of
Eq.(\ref{d1}) in the region of intermediate $q$.  Using  (\ref{d1}) we can
write:

\begin{eqnarray}
& &\sum_{\bf q}\sqrt{G_{0}\otimes G_{0}}exp\left\{-\frac{1}{2W^{2}G_{0}\otimes G_{0}}\right\}=\nonumber \\
&=&\frac{1}{2\pi ^{2}}\int\limits_{0}^{\pi\sqrt{|E|}}dqq^{2}\sqrt{\frac{1}{8\pi\sqrt{|E|}}}
exp\left\{-\frac{4\pi\sqrt{|E|}}{W^{2}}\right\}+
\frac{1}{2\pi ^{2}}\int\limits_{\pi\sqrt{|E|}}^{\infty}dqq^{2}\sqrt{\frac{1}{8q}}
exp\left\{-\frac{4q}{W^{2}}\right\}\approx\nonumber \\
&\approx &\frac{\sqrt{\pi}}{12\sqrt{2}}|E|^{\frac{5}{4}}\{1+O(W^{2})\}
exp\left\{-\frac{4\pi\sqrt{|E|}}{W^{2}}\right\}
\label{d2}
\end{eqnarray}
In fact the main contribution here is from the first integral. Then:

\begin{equation}
\left|\sum_{\bf p}G_{0}(p)\right|=
\frac{1}{\pi ^{2}}\int\limits_{0}^{p_{0}}dp\frac{p^{2}}{|E|+p^{2}}
\approx\frac{1}{2\pi ^{2}}\left(p_{0}-\frac{\pi}{2}\sqrt{|E|}\right)
\approx\frac{p_{0}}{2\pi ^{2}}
\label{d3}
\end{equation}
Putting all these expressions into Eq.(\ref{49}) and restoring $2m$, 
we obtain (\ref{52}).

\newpage

{\sl Table I}

\flushleft

\vskip 0.5cm

$
\begin{tabular}{|c|c|c|c|c|}
\hline
N & $\Gamma _N=a_N$ & b$_N=a_N/(2N+1)!!$ & $\Sigma _N=a_{N-1}$ & U$%
_N=(2N-1)a_{N-1}$ \\ \hline
1 & 1 & 0.3333 & 1 & 1 \\ \hline
2 & 4 & 0.2667 & 1 & 3 \\ \hline
3 & 27 & 0.2571 & 4 & 20 \\ \hline
4 & 248 & 0.2624 & 27 & 189 \\ \hline
5 & 2830 & 0.2722 & 248 & 2232 \\ \hline
6 & 38232 & 0.2829 & 2830 & 3130 \\ \hline
7 & 593859 & 0.2930 & 38232 & 497016 \\ \hline
8 & 10401712 & 0.3019 & 593859 & 8907885 \\ \hline
9 & 202601898 & 0.3158 & 10401712 & 176829104 \\ \hline
10 & 4342263000 & 0.3211 & 202601898 & 3849436062 \\ \hline
$N\gg 1$ & $\frac 1e[1-\frac 5{4N}](2N+1)!!$ & $\frac 1e[1-\frac 5{4N}]$ 
& $\frac 1e[1-\frac 5{4N}](2N-1)!!$ & $\frac 1e[1-\frac 9{4N}](2N+1)!!$ \\ 
\hline
\end{tabular}
$

\endflushleft

\newpage
\setlength{\textwidth}{15.0cm}
\begin{center}
{\bf Figure Captions.}
\end{center}
\vskip 0.5cm

Fig.1. Diagrammatic expansion for the averaged one-particle Green's
function (a) and self-energy part (b). Dashed line corresponds to the
quadratic correlator of the random field. $G_{0}$ -- free particle Green's
function.

\vskip 0.3cm

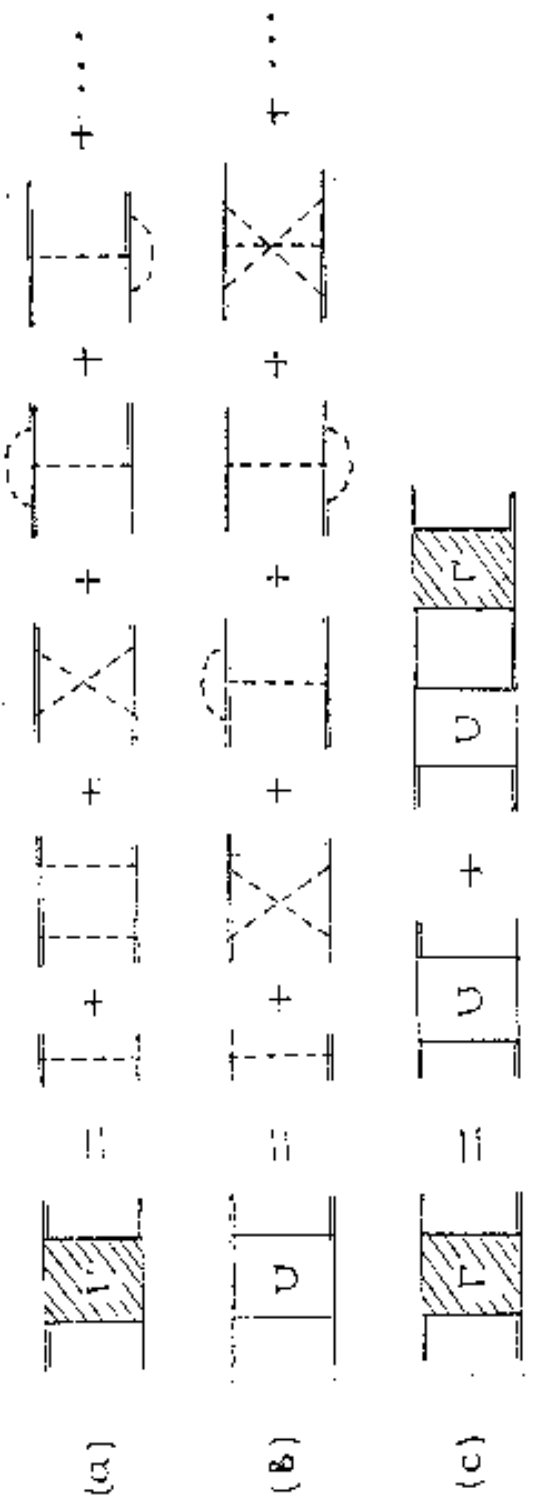
Fig.2. Diagrammatic expansion for the full vertex-part $\Gamma$ (a), for the
irreducible vertex-part $U$ (b) and Bethe-Salpeter equation, connecting
$\Gamma$ and $U$ (c).

\vskip 0.3cm

Fig.3. Equation connecting the self-energy part with the full vertex-part.

\vskip 0.3cm

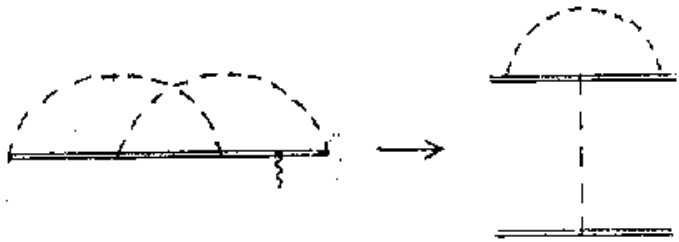
Fig.4. The cut of any of $2N-1$ internal Green's function lines in the
``skeleton'' diagram of  $N$-th order for the self-energy part produces the
appropriate diagram for $U$.

\vskip 0.3cm

Fig.5. The behavior of $b_{n}$ with the growth of $n$. Dots represent the
values of $b_{n}$, obtained from the recurrence relation (\ref{28}), 
the curve represents the asymptotic dependence 
$\frac{1}{e}\left(1-\frac{5}{4}\frac{1}{n}\right)$, dashed line --- the 
asymptote  $\frac{1}{e}$.

\vskip 0.3cm

Fig.6. (a) -- ``basic'' diagram, used to construct the approximation for the
self-energy part,
(b) -- ``rotated maximally crossed'' diagrams produce the ``ladder'' in
case of time reversal invariant system.

\vskip 0.3cm

Fig.7. Density of states in one-dimensional system for the different values
of the averaged square of the random field 
$\frac{W^{2}(2m)^{\frac{1}{2}}}{E_{0}^{\frac{3}{2}}}$: 1. 0.25; 2. 2; 3. 16.
Continuous curves --- the exact solution,\ dashed lines ---
self-consistent Born approximation (\ref{57}). Energy is in units of $E_{0}$, 
density of states -- in units of $\frac{\sqrt{2m}}{\sqrt{E_{0}}}$,\  
$E_{0}$ -- some arbitrary unit of energy.

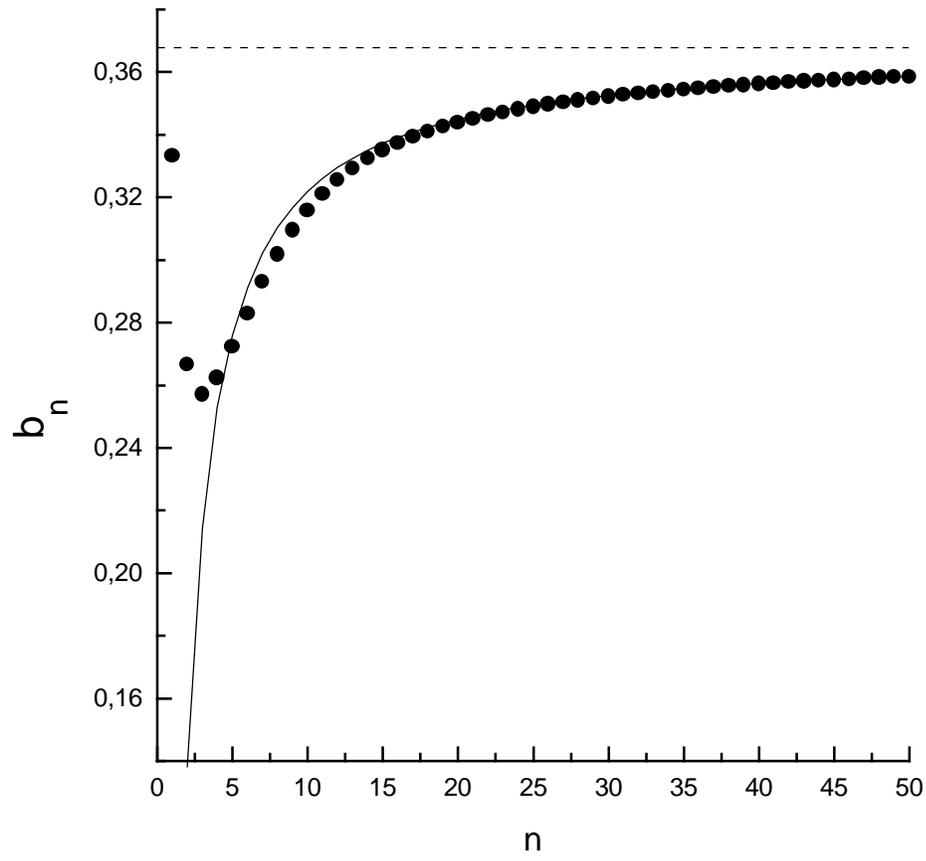
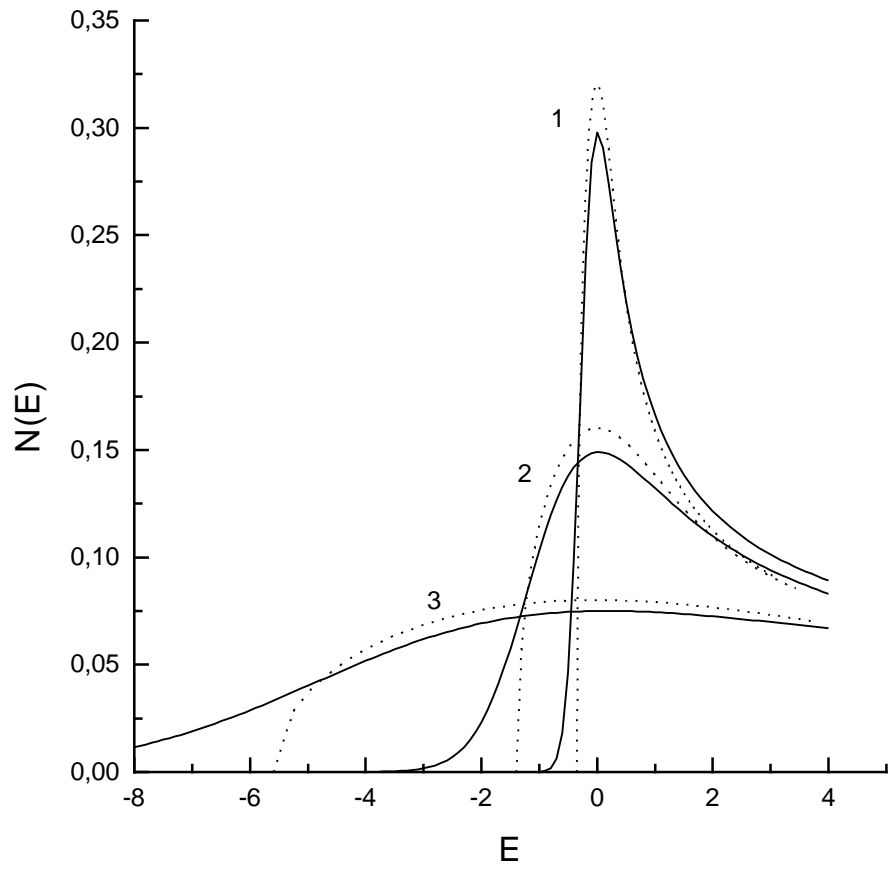
\end{document}